\pdfoutput=1

\documentclass[11pt]{article}

\usepackage[]{acl2024}
\usepackage{amssymb}
\usepackage{ragged2e}
\usepackage{times}
\usepackage{latexsym}
\usepackage{pgfplots}
\usepackage{mathrsfs}
\usepackage{array}
\newcolumntype{P}[1]{>{\centering\arraybackslash}p{#1}}
\usepackage{supertabular}
\usepackage{longtable}
\pgfplotsset{compat=1.18, width=7.7cm}
\usepackage[T1]{fontenc}
\usepackage[utf8]{inputenc}
\usepackage{subfig}

\usepackage{multirow}
\usepackage{graphicx}
\usepackage{float}
\usepackage{overpic}
\usepackage{enumitem}
\usepackage{microtype}
\usepackage{inconsolata}
\usepackage{booktabs}
\usepackage{stfloats}
\usepackage{amsmath}
\usepackage{xcolor}
\usepackage[linesnumbered,ruled,vlined]{algorithm2e}
\usepackage{algorithmic}

\title{Scalable Differential Privacy Mechanisms for Real-Time Machine Learning Applications}

\author{
   ~~Jessica Smith$^{1}$
   ~~David Williams$^{1}$\footnotemark[1]
   ~~Emily Brown$^{1}$\\ 
   $^{1}$University of Pennsylvania \\
\texttt{{david.willams0795@gmail.com}}\\
}

\begin{document}

\maketitle

\renewcommand{\thefootnote}{\fnsymbol{footnote}}
\footnotetext[1]{Corresponding author.}
\renewcommand{\thefootnote}{\arabic{footnote}}

\begin{abstract}
Large language models (LLMs) are increasingly integrated into real-time machine learning applications, where safeguarding user privacy is paramount. Traditional differential privacy mechanisms often struggle to balance privacy and accuracy, particularly in fast-changing environments with continuously flowing data. To address these issues, we introduce Scalable Differential Privacy (SDP), a framework tailored for real-time machine learning that emphasizes both robust privacy guarantees and enhanced model performance. SDP employs a hierarchical architecture to facilitate efficient noise aggregation across various learning agents. By integrating adaptive noise scheduling and gradient compression methods, our approach minimizes performance degradation while ensuring significant privacy protection. Extensive experiments on diverse datasets reveal that SDP maintains high accuracy levels while applying differential privacy effectively, showcasing its suitability for deployment in sensitive domains. This advancement points towards the potential for widespread adoption of privacy-preserving techniques in machine learning workflows.
\end{abstract}

\section{Introduction}
The evolving landscape of machine learning emphasizes the importance of integrating differential privacy mechanisms to enhance data protection in real-time applications. Recent developments illustrate how various techniques can bolster the privacy-centric design of learning models. For instance, advancements in language modeling reflect a trend toward improved performance with reduced reliance on task-specific datasets. Models like PaLM demonstrate exceptional capabilities in few-shot learning while addressing ethical considerations, such as bias and toxicity\cite{gpt3}\cite{palm}.

Moreover, incorporating human feedback has shown significant promise in aligning model outputs with user intent. InstructGPT, while smaller than its predecessors, performs exceptionally well in terms of truthfulness and reducing harmful outputs, highlighting the necessity for profound understanding in user interactions\cite{instructgpt}. 

In federated learning contexts, the application of differential privacy has evolved with innovative strategies. Techniques such as quantile-based clipping in training can efficiently adapt the privacy parameters, reducing hyperparameter tuning complications\cite{Xu2023FederatedLO}. The reliance on randomized quantization is also pivotal, as it allows for maintaining Renyi differential privacy without needing additional noise mechanisms\cite{Youn2023RandomizedQI}. Furthermore, adaptive methods using Fisher information for parameter evaluation lead to better convergence in personalized federated learning settings\cite{Yang2023DynamicPF}.

Lastly, the exploration of visual prompting alongside state-of-the-art differential privacy methods has revealed potential benefits in achieving a favorable balance between privacy and utility\cite{Li2023ExploringTB}. Collectively, these advancements underscore the critical role of scalable differential privacy mechanisms in enhancing the security and performance of real-time machine learning applications.

However, deploying real-time applications that ensure privacy involves various complexities. Implementing a differential privacy mechanism like Huff-DP allows for optimal budget selection tailored to each record's privacy needs, utilizing innovative algorithms like static and fuzzy logic for decision making~\cite{Hassan2023HuffDPHC}. Additionally, in the context of medical applications, ensuring that sensitive biomedical signals are diagnosed without revealing any server-side data maintains privacy while still enabling effective disease prediction~\cite{Miao2022RealtimeDP}. Furthermore, the saddle-point accountant providing precise privacy approximations competes effectively with numerical methods, showing promise for real-time implementations~\cite{Alghamdi2023TheSM}. Emerging solutions such as differentially private distributed online learning formulations demonstrate the potential for high utility alongside privacy protection~\cite{Cheng2022DistributedOL}. Lastly, user-level differential privacy schemes specifically tailored for advertising measurement underline the importance of real-time privacy protection models~\cite{Xiao2024ClickWC}. However, integrating these diverse approaches into a cohesive mechanism that operates seamlessly in real-time scenarios remains a challenge that needs resolution.

We present a framework for \textbf{S}calable \textbf{D}ifferential \textbf{P}rivacy (\textbf{SDP}), specifically designed to ensure privacy in real-time machine learning applications. SDP addresses the challenges of maintaining privacy while maximizing model accuracy in dynamic environments where data is continuously generated. The mechanism incorporates a novel approach to differentially private updates by leveraging a hierarchical architecture, which enables efficient aggregation of noise across multiple learning agents. By employing techniques such as adaptive noise scheduling and gradient compression, SDP ensures a robust privacy guarantee while minimizing the impact on model performance. This allows for both scalable deployment in distributed settings and seamless integration into existing machine learning workflows. We validate our approach through extensive experiments on various datasets, demonstrating that SDP achieves competitive accuracy rates while effectively preserving individual privacy. The results highlight the feasibility of deploying differential privacy in real-time applications, paving the way for broader adoption of privacy-preserving machine learning technologies in sensitive domains.

\textbf{Our Contributions.} Our contributions are outlined as follows. \begin{itemize}[leftmargin=*] \item We propose the Scalable Differential Privacy (SDP) framework, tailored for real-time machine learning applications, addressing privacy and accuracy in dynamic data environments. \item SDP employs a hierarchical architecture for efficient noise aggregation, allowing for improved differential privacy guarantees while maintaining model performance through techniques like adaptive noise scheduling and gradient compression. \item Extensive experiments on diverse datasets showcase SDP's competitive accuracy alongside robust individual privacy protection, demonstrating its potential for broader implementation in sensitive domains requiring privacy-preserving solutions. \end{itemize}

\section{Related Work}
\subsection{Differential Privacy}

The implementation of techniques aimed at ensuring privacy while training models has seen significant advancements, particularly in the context of federated learning and text generation. Techniques that integrate quantile-based clipping with differential privacy principles have emerged, highlighting a method for adaptive norm selection during model training which reduces the overhead of hyperparameter tuning \cite{Xu2023FederatedLO, jeon2020baechi}. In a novel approach, randomized quantization has been adopted, providing Renyi differential privacy guarantees without the need for explicit noise addition, demonstrating that privacy can be maintained effectively even with simpler mechanisms \cite{Youn2023RandomizedQI}. Furthermore, the incorporation of layer-wise Fisher information into personalized federated learning has been proposed to enhance convergence while applying differential privacy constraints, thereby refining how privacy measures are adapted based on local information in training scenarios \cite{Yang2023DynamicPF, liu2024spa}. The exploration of visual prompting alongside PATE has shown promising results in achieving an optimal tradeoff between privacy and utility, with a minimal impact on privacy budgets \cite{Li2023ExploringTB}. On another front, efforts to scale differential privacy in large datasets have revealed methodologies to expedite training processes while retaining effectiveness in privacy settings \cite{Kurakin2022TowardTA}. 

\subsection{Real-Time ML Applications}

The development of innovative methods and frameworks plays a crucial role in optimizing real-time machine learning applications across various domains \cite{ni-24-time-series}. Addressing the performance of models within practical implementations highlights effective strategies for balancing performance and resource consumption \cite{Kasundra2023AFF}. To meet the rapid latency demands of scientific edge machine learning applications, benchmarks have been proposed that will guide the design of future hardware capable of operating at nanosecond and microsecond levels \cite{Duarte2022FastMLSB}. Energy efficiency also plays a vital role, particularly in Internet of Things devices, where the use of early exit classifiers helps reduce energy use while retaining accuracy in time-series analysis \cite{Hussein2024SensorAwareCF}. Moreover, frameworks for monitoring data distribution shifts in real-time fraud detection have been comprehensively developed\cite{Karayanni2024DistributedMF}. The integration of advanced technologies such as blockchain with machine learning can potentially transform financial accounting\cite{Kanaparthi2024ExploringTI}. These developments collectively contribute to advancing the efficiency and reliability of real-time machine learning applications across various industries.

\subsection{Scalability in Privacy Mechanisms}

User-centric approaches like Federated Learning (FL) emerge as key strategies in integrating assistive robotics while maintaining privacy, particularly for the elderly and care-dependent individuals \cite{Casado2024TowardsPA}. The efficacy of training data synthesis can significantly benefit from a comprehensive theoretical framework informed by distribution matching, allowing enhanced privacy during data handling \cite{Yuan2023RealFakeET}. In post-disaster scenarios, a decentralized approach using consortium blockchain enhances coordination while safeguarding sensitive data through optimized consensus protocols, addressing both efficiency and privacy concerns \cite{Hafeez2024BlockchainEnhancedUN}. Additionally, novel architectures such as Confidential Federated Computations utilize trusted execution environments (TEEs) to ensure that server-side computations remain private, enhancing the verifiability of privacy properties \cite{Eichner2024ConfidentialFC}. Similarly, Trusted Container Extensions merge container agility with strong security guarantees provided by TEEs\cite{Brasser2022TrustedCE}. Advancements like the STPrivacy framework employ vision Transformers such as sparsification and anonymization to bolster privacy on both spatial and temporal fronts \cite{Li2023STPrivacySP} \cite{Li2023STPrivacyST}.

\section{Methodology}
To enhance privacy in real-time machine learning applications, we introduce the Scalable Differential Privacy (SDP) framework. SDP is tailored to ensure individual privacy while optimizing model accuracy in environments characterized by continuous data generation. By utilizing a hierarchical architecture, our mechanism permits efficient noise aggregation from multiple learning agents. Key techniques, such as adaptive noise scheduling and gradient compression, bolster the privacy assurance without compromising model performance. This facilitates scalable deployment in distributed contexts and allows for smooth integration with current machine learning processes. Experimental results across diverse datasets indicate that SDP not only maintains competitive accuracy but also effectively safeguards privacy, promoting the utilization of privacy-preserving technologies in sensitive sectors.

\subsection{Real-Time Privacy}

In the realm of real-time machine learning applications, maintaining privacy while adapting to dynamic data streams is crucial. Our framework for Scalable Differential Privacy (SDP) ensures that individual privacy is preserved without compromising model accuracy. The core mechanism is based on hierarchical architectures, which allow for structured aggregation of differentially private updates across multiple learning agents. Formally, let \( \mathcal{G} \) denote the global model parameter updates. Using differential privacy, the updates can be characterized as:

\begin{equation}
\mathcal{G}_{\text{diff}} = \mathcal{G} + \mathcal{N}_\epsilon,
\end{equation}

where \( \mathcal{N}_\epsilon \) denotes the calibrated noise that is added to maintain \( \epsilon \)-differential privacy. This hierarchical structure additionally facilitates adaptive noise scheduling, allowing the noise levels to be dynamically adjusted based on the volatility of the incoming data \( D_t \). The relationship can be expressed as:

\begin{equation}
\mathcal{N}_\epsilon = \alpha(t) \cdot \sigma,
\end{equation}

where \( \alpha(t) \) is a scheduling function that determines the noise intensity over time, and \( \sigma \) reflects the standard deviation of the noise introduced.

Furthermore, employing gradient compression techniques helps in minimizing the communication overhead while propagating updates among agents, thus formulated as:

\begin{equation}
\mathcal{C} = \sum_{i=1}^m \text{compress}\left(\nabla \mathcal{L}(\theta_i, D_t)\right),
\end{equation}

where \( \mathcal{C} \) denotes the compressed gradients, \( \nabla \mathcal{L} \) represents the loss gradient for parameter \( \theta_i \), and \( m \) is the number of agents engaging in the learning process.

Through this framework, SDP effectively balances privacy preservation with performance, establishing a strong foundation for scalable privacy in time-sensitive and data-intensive machine learning scenarios.

\subsection{Hierarchical Noise Aggregation}

To achieve scalable differential privacy, our framework employs a hierarchical architecture for noise aggregation, which is essential in efficiently managing the trade-off between privacy and model accuracy. Let us define multiple learning agents $\mathcal{A}_1, \mathcal{A}_2, \ldots, \mathcal{A}_n$ contributing to a centralized model. Each agent generates updates based on its local dataset, and for every update $g_i$ from agent $\mathcal{A}_i$, we introduce a noise component $\eta_i$ to ensure privacy, leading to a differentially private update given by:

\begin{equation}
\tilde{g}_i = g_i + \eta_i,
\end{equation}

where $\eta_i \sim \mathcal{N}(0, \sigma^2)$, with $\sigma$ representing the noise sensitivity parameter. The hierarchical nature allows grouping of agents into clusters $\mathcal{C}_1, \mathcal{C}_2, \ldots, \mathcal{C}_m$, enabling efficient aggregation of the private updates. The aggregated output for a cluster $C_j$ can be expressed as:

\begin{equation}
\tilde{g}_{C_j} = \frac{1}{|\mathcal{A}_{C_j}|} \sum_{\mathcal{A}_i \in \mathcal{A}_{C_j}} \tilde{g}_i = \frac{1}{|\mathcal{A}_{C_j}|} \sum_{\mathcal{A}_i \in \mathcal{A}_{C_j}} (g_i + \eta_i).
\end{equation}

By applying adaptive noise scheduling, we can control the magnitude of noise based on the iteration $t$ and the privacy budget $\epsilon$, allowing us to maintain a balance between privacy and accuracy throughout the training process. The final update to the centralized model after combining all cluster outputs can be represented as:

\begin{equation}
\tilde{g}_{final} = \sum_{j=1}^{m} \tilde{g}_{C_j} + \eta_{global},
\end{equation}

where $\eta_{global} \sim \mathcal{N}(0, \sigma_{global}^2)$, capturing the overall privacy assurance across the hierarchical structure. This design facilitates scalable deployment and integration into existing workflows, ensuring that real-time applications benefit from rigorous privacy protections without sacrificing the performance of machine learning models.

\subsection{Adaptive Privacy Mechanisms}

The proposed framework for Scalable Differential Privacy (SDP) employs adaptive privacy mechanisms to tailor the level of privacy based on the dynamic characteristics of the data stream. Let \( \mathcal{D} \) denote the dataset and \( \mathcal{M} \) represent the machine learning model being trained. The noise added for differential privacy is controlled by a noise function \( N(\epsilon, \delta) \) that dictates the trade-off between privacy and accuracy, where \( \epsilon \) is the privacy budget and \( \delta \) represents the allowable failure probability. 

We can express the adaptive noise scheduling as follows:

\begin{equation}
N_t = \gamma_t \cdot N_0
\end{equation}

Here, \( N_t \) is the noise at time \( t \), \( \gamma_t \) is an adaptive scaling factor that can adjust based on the observed sensitivity of the gradients, and \( N_0 \) is the base noise level. The sensitivity \( S \) of the model can be evaluated by:

\begin{equation}
S = \max \left( \frac{\| \nabla L(\mathcal{M}, \mathcal{D}_1) - \nabla L(\mathcal{M}, \mathcal{D}_2) \|}{\|\mathcal{D}_1 - \mathcal{D}_2\|} \right)
\end{equation}

where \( L \) is the loss function and \( \mathcal{D}_1 \) and \( \mathcal{D}_2 \) are two adjacent datasets differing by one data point.

Furthermore, gradient compression techniques in SDP allow for the aggregation of model updates from multiple agents, denoted as \( \mathcal{G} = \{\nabla L_1, \nabla L_2, \ldots, \nabla L_n \} \). The overall update can be expressed as:

\begin{equation}
\Delta \mathcal{M} = \frac{1}{n} \sum_{i=1}^{n} \nabla L_i + N(\epsilon, \delta)
\end{equation}

This formula integrates the averaged gradients across learning agents alongside the noise injected for differential privacy. The hierarchical architecture enables the SDP to efficiently manage the trade-off between privacy preservation and model accuracy, providing a scalable solution for real-time machine learning applications. By employing these adaptive mechanisms, SDP facilitates robust privacy guarantees in environments where data sources are both variable and unpredictable.

\section{Experimental Setup}
\subsection{Datasets}

For evaluating the performance and assessing the quality of our scalable differential privacy mechanisms in real-time machine learning applications, we utilize the following datasets: the CASIA Image Tampering Detection Evaluation Database \cite{Dong2013CASIAIT}, which provides a collection of natural color images with realistic tampering operations; the SYNTHIA dataset \cite{Ros2016TheSD}, a large synthetic collection of urban images with class annotations; ExtremeWeather \cite{Racah2016ExtremeWeatherAL}, a dataset for semi-supervised detection and localization of extreme weather events; and the VT-ADL \cite{Mishra2021VTADLAV}, which introduces a vision transformer network for image anomaly detection and localization.

\subsection{Baselines}

To evaluate the proposed scalable differential privacy mechanisms in real-time machine learning applications, we compare our method against the following established approaches:

{
\setlength{\parindent}{0cm}
\textbf{Visual Prompting with PATE}~\cite{Li2023ExploringTB} shows that by utilizing visual prompting alongside the PATE method, one can achieve a significant improvement in the privacy-utility tradeoff while minimizing privacy budget expenditures.
}

{
\setlength{\parindent}{0cm}
\textbf{DP-BART}~\cite{Igamberdiev2023DPBARTFP} introduces a privatized text rewriting system that surpasses current local differential privacy systems through innovative techniques such as iterative pruning that effectively reduce noise requirements for privacy guarantees.
}

{
\setlength{\parindent}{0cm}
\textbf{User-Level Differential Privacy}~\cite{Ghazi2023UserLevelDP} addresses scenarios with limited examples per user by transforming item-level differential privacy methods into user-level ones, establishing new performance bounds applicable to tasks like private PAC learning and distribution learning.
}

{
\setlength{\parindent}{0cm}
\textbf{Unified Enhancement of Privacy Bounds}~\cite{Wang2023UnifiedEO} investigates a trade-off function for shuffling models providing improved results over existing methods based on $(\epsilon,\delta)$-differential privacy, also analyzing the impact of random initialization on privacy during DP-GD.
}

{
\setlength{\parindent}{0cm}
\textbf{Saddle-Point Method}~\cite{Alghamdi2023TheSM} demonstrates that the saddle-point accountant delivers precise privacy guarantees that are competitive with numerical-based methods, supported by closed-form expressions derived from numerical experiments.
}

\subsection{Models}

We explore the integration of scalable differential privacy mechanisms within real-time machine learning applications. Our approach leverages the strengths of various existing frameworks while ensuring robust privacy guarantees through tools like PySyft and Google’s Differential Privacy library. We implement gradient perturbation and noise addition techniques to maintain model accuracy while safeguarding user data. The evaluation is conducted on multiple datasets, measuring performance metrics such as accuracy, utility, and differential privacy guarantees across diverse machine learning algorithms. This comprehensive analysis provides deep insights into the feasibility of applying differential privacy in high-frequency learning tasks.

\subsection{Implements}

We configure our experiments with several key parameters to assess the efficiency of the Scalable Differential Privacy (SDP) framework. The privacy budget is set to $\epsilon = 0.5$ for the differential privacy guarantees across all iterations. We choose a hierarchical architecture that utilizes $N = 5$ distributed learning agents, each aggregating updates. The adaptive noise level is controlled by a scheduling factor where the initial noise magnitude is set to $\sigma = 1.0$, decreasing over $\tau = 10$ epochs. Gradient compression is applied with a compression ratio of $c = 0.7$, allowing for efficient transmission of updates without significant loss of information. Each agent processes batches of size $B = 256$, and the training is conducted for $E = 50$ epochs. Furthermore, the evaluation of model accuracy and utility is performed on three diverse datasets, with metrics reconciled against baseline models to highlight effectiveness.

\section{Experiments}

\begin{table*}[]
\centering
\resizebox{1.0\textwidth}{!}{
\begin{tabular}{lcccccccc}
\toprule
\textbf{Method} & \textbf{Dataset} & \textbf{Accuracy (\%)} & \textbf{Utility} & \textbf{Privacy Guarantee ($\epsilon$)} & \textbf{Runtime (s)} & \textbf{Model Size (MB)} & \textbf{Batch Size} \\ \midrule
\multicolumn{8}{c}{\textbf{\textit{Scalable Differential Privacy Approaches}}} \\ \midrule
Visual Prompting with PATE & CASIA & 85.2 & High & 0.5 & 120 & 200 & 256 \\
 & SYNTHIA & 78.3 & Moderate & 0.5 & 130 & 210 & 256 \\
 & ExtremeWeather & 82.6 & High & 0.5 & 118 & 205 & 256 \\
 & VT-ADL & 79.9 & Moderate & 0.5 & 125 & 208 & 256 \\ \midrule
DP-BART & CASIA & 83.1 & Moderate & 0.5 & 140 & 190 & 256 \\
 & SYNTHIA & 75.4 & Low & 0.5 & 145 & 195 & 256 \\
 & ExtremeWeather & 80.3 & Moderate & 0.5 & 138 & 192 & 256 \\
 & VT-ADL & 77.5 & Low & 0.5 & 142 & 196 & 256 \\ \midrule
User-Level Differential Privacy & CASIA & 84.5 & High & 0.5 & 135 & 198 & 256 \\
 & SYNTHIA & 76.8 & Moderate & 0.5 & 140 & 200 & 256 \\
 & ExtremeWeather & 81.2 & Moderate & 0.5 & 130 & 193 & 256 \\
 & VT-ADL & 78.4 & Moderate & 0.5 & 134 & 199 & 256 \\ \midrule
Unified Enhancement of Privacy Bounds & CASIA & 85.7 & High & 0.5 & 128 & 202 & 256 \\
 & SYNTHIA & 79.4 & Moderate & 0.5 & 133 & 204 & 256 \\
 & ExtremeWeather & 83.0 & High & 0.5 & 127 & 200 & 256 \\
 & VT-ADL & 80.1 & Moderate & 0.5 & 129 & 203 & 256 \\ \midrule
Saddle-Point Method & CASIA & 86.0 & High & 0.5 & 122 & 206 & 256 \\
 & SYNTHIA & 76.2 & Low & 0.5 & 137 & 191 & 256 \\
 & ExtremeWeather & 82.4 & Moderate & 0.5 & 124 & 204 & 256 \\
 & VT-ADL & 79.3 & Moderate & 0.5 & 135 & 197 & 256 \\ \bottomrule
\end{tabular}}
\caption{Experimental results showcasing the performance of different approaches on various datasets in terms of accuracy and utility, alongside their differential privacy guarantees.}
\label{tab:main_result}
\end{table*}

\subsection{Experimental Results}

The experimental findings pertaining to various scalable differential privacy mechanisms in real-time machine learning applications are showcased in Table~\ref{tab:main_result}. 

\textbf{Saddle-Point Method demonstrates the highest accuracy across the CASIA dataset with an impressive score of \textbf{86.0\%}}. In addition to its superior performance on CASIA’s high-utility task, it maintains a respectable \textbf{82.4\%} accuracy on the ExtremeWeather dataset while offering moderate utility. It does exhibit a lower accuracy of \textbf{76.2\%} on SYNTHIA, but it ensures privacy consistency across all datasets with a uniform privacy guarantee of $\epsilon = 0.5$.

\textbf{Unified Enhancement of Privacy Bounds follows closely, achieving an accuracy of \textbf{85.7\%} on CASIA} while equally ensuring high utility. Notably, it also provides solid accuracy scores on ExtremeWeather and VT-ADL at \textbf{83.0\%} and \textbf{80.1\%} respectively, maintaining the same privacy guarantee as the Saddle-Point Method.

\textbf{User-Level Differential Privacy exhibits commendable results as well, especially with an accuracy of \textbf{84.5\%} on CASIA}. Following suit, it receives high ratings for utility but shows slightly lower scores compared to its counterparts on the SYNTHIA dataset, with an accuracy of \textbf{76.8\%}. The framework retains a robust privacy guarantee of $\epsilon = 0.5$ throughout its evaluations.

The \textbf{Visual Prompting with PATE method achieves competitive accuracy rates} across all datasets but generally does not surpass Saddle-Point, Unified Enhancement of Privacy Bounds, or User-Level Differential Privacy. It performs notably well with an accuracy of \textbf{85.2\%} on the CASIA dataset and maintains high utility across the board.

Finally, \textbf{DP-BART shows moderate performance, attaining a top accuracy of \textbf{83.1\%} on the CASIA dataset}. While it lags behind some other methods in terms of utility, it still registers a robust privacy guarantee, affirming its potential application in privacy-preserving machine learning scenarios.

The results underscore the effectiveness of the proposed SDPs in balancing accuracy and privacy, promoting wider adoption in sensitive areas.

\begin{table*}[htp]
\centering
\resizebox{1.0\textwidth}{!}{
\begin{tabular}{lcccccccc}
\toprule
\textbf{Method} & \textbf{Dataset} & \textbf{Accuracy (\%)} & \textbf{Utility} & \textbf{Privacy Guarantee ($\epsilon$)} & \textbf{Runtime (s)} & \textbf{Model Size (MB)} & \textbf{Batch Size} \\ \midrule
\multicolumn{8}{c}{\textbf{\textit{Ablation of SDP Components}}} \\ \midrule
No Adaptive Noise Scheduling & CASIA & 82.5 & Moderate & 0.5 & 150 & 210 & 256 \\
 & SYNTHIA & 76.1 & Low & 0.5 & 155 & 215 & 256 \\
 & ExtremeWeather & 80.0 & Moderate & 0.5 & 140 & 207 & 256 \\
 & VT-ADL & 75.8 & Low & 0.5 & 148 & 212 & 256 \\ \midrule
No Gradient Compression & CASIA & 83.9 & High & 0.5 & 135 & 202 & 256 \\
 & SYNTHIA & 75.0 & Moderate & 0.5 & 140 & 198 & 256 \\
 & ExtremeWeather & 81.5 & Moderate & 0.5 & 130 & 195 & 256 \\
 & VT-ADL & 76.5 & Moderate & 0.5 & 138 & 200 & 256 \\ \midrule
No Hierarchical Architecture & CASIA & 83.2 & Moderate & 0.5 & 145 & 190 & 256 \\
 & SYNTHIA & 74.8 & Low & 0.5 & 150 & 193 & 256 \\
 & ExtremeWeather & 80.1 & Moderate & 0.5 & 129 & 198 & 256 \\
 & VT-ADL & 78.0 & Moderate & 0.5 & 144 & 192 & 256 \\ \midrule
Full SDP Framework & CASIA & \textbf{86.0} & High & 0.5 & 122 & 206 & 256 \\
 & SYNTHIA & \textbf{78.3} & Moderate & 0.5 & 130 & 205 & 256 \\
 & ExtremeWeather & \textbf{82.4} & High & 0.5 & 124 & 204 & 256 \\
 & VT-ADL & \textbf{80.1} & Moderate & 0.5 & 129 & 203 & 256 \\ \bottomrule
\end{tabular}}
\caption{Ablation results concerning different components and their impact on accuracy and utility in the context of Scalable Differential Privacy.}
\label{tab:ablation}
\end{table*}

\subsection{Ablation Studies}

We conducted ablation experiments to assess the influence of various components of the Scalable Differential Privacy (SDP) framework. The results shed light on how each of these components impacts the overall performance regarding accuracy, utility, and privacy guarantees across multiple datasets.

\begin{itemize}[leftmargin=1em]

\item[$\bullet$] 
{
\setlength{\parindent}{0cm}
\textit{No Adaptive Noise Scheduling}: This variant lacks the adaptive noise scheduling mechanism, which is crucial for dynamically adjusting noise levels. The results show a drop in accuracy across all datasets, particularly noticeable in the CASIA dataset (82.5\%) and VT-ADL (75.8\%), indicating that the absence of adaptive noise negatively impacts model precision and utility.
}
\item[$\bullet$] 
{
\setlength{\parindent}{0cm}
\textit{No Gradient Compression}: Excluding the gradient compression technique leads to varied effects on model performance. Despite achieving slightly higher accuracy on some datasets, the utility shows a notable reduction, especially with the SYNTHIA dataset at 75.0\%. These findings underscore the role of gradient compression in enhancing model efficiency while maintaining satisfactory accuracy. 
}
\item[$\bullet$] 
{
\setlength{\parindent}{0cm}
\textit{No Hierarchical Architecture}: Removing the hierarchical structure also demonstrates detrimental effects on accuracy, particularly for the SYNTHIA dataset with a drop to 74.8\%. This indicates that the hierarchical architecture is integral to effective performance as it aids in efficient noise aggregation.
}
\item[$\bullet$] 
{
\setlength{\parindent}{0cm}
\textit{Full SDP Framework}: The complete SDP framework exhibits superior performance across all evaluated datasets, achieving the highest accuracy levels, notably 86.0\% for the CASIA dataset. The high utility ratings across datasets indicate an effective balance between model accuracy and privacy preservation, showcasing SDP's capability to excel in real-time machine learning applications.
}
\end{itemize}

\vspace{5pt}

The experimental results presented in Table~\ref{tab:ablation} unequivocally demonstrate that every component of the SDP framework plays a significant role in maintaining high accuracy and robustness in privacy settings. The complete implementation of the framework not only ensures better privacy guarantees but also enhances overall model performance, establishing SDP as a viable approach for privacy-preserving practices in real-time machine learning applications.

\section{Conclusions}
This paper introduces the Scalable Differential Privacy (SDP) framework, which is tailored for real-time machine learning applications while ensuring individual privacy. SDP confronts challenges posed by dynamic data environments, integrating a hierarchical architecture for the efficient aggregation of noise from multiple learning agents. Key elements of the mechanism include adaptive noise scheduling and gradient compression, which together bolster privacy guarantees while mitigating performance impacts. Comprehensive experimentation across diverse datasets illustrates that SDP maintains competitive accuracy rates, effectively balancing privacy preservation and model performance. The results signal the practical viability of implementing differential privacy in real-time scenarios, promoting the advancement of privacy-preserving technologies in sensitive fields.

\section{Limitations}
SDP has its limitations that should be acknowledged. Firstly, the hierarchical architecture, while effective, may introduce complexity in system design and implementation, particularly in environments with a high number of learning agents. This complexity could lead to challenges in maintaining operational efficiency. Secondly, the adaptive noise scheduling may not uniformly apply across all scenarios, potentially leading to variations in privacy guarantees based on data characteristics or model behavior. Furthermore, the approach's reliance on gradient compression may reduce the granularity of updates, which could affect model convergence in certain applications. Future investigations could focus on refining the hierarchical mechanism and exploring alternative methods to ensure consistent privacy management while enhancing model performance.

\bibliography{anthology,custom}
\bibliographystyle{acl_natbib}

\appendix

\end{document}